\documentclass{article}
\usepackage{spconf,amsmath,graphicx}
\usepackage{hyperref}
\usepackage{amsfonts} 
\usepackage{multirow}
\usepackage{colortbl}
\usepackage{verbatim}
\usepackage[table,xcdraw]{xcolor}

\definecolor{Gray}{gray}{0.85}
\definecolor{LightCyan}{rgb}{0.88,1,1}
\newcolumntype{a}{>{\columncolor{Gray}}c}
\newcolumntype{i}{>{\columncolor{white}}c}

\title{Less peaky and more accurate CTC forced alignment by label priors}
\name{
\vspace{-2pt}
\begin{tabular}{c} 
Ruizhe Huang$^{1\dagger}$, Xiaohui Zhang$^2$, Zhaoheng Ni$^2$, Li Sun$^{5\dagger}$, Moto Hira$^2$, Jeff Hwang$^2$, Vimal Manohar$^2$,\\ Vineel Pratap$^2$, Matthew Wiesner$^1$, Shinji Watanabe$^3$, Daniel Povey$^4$, Sanjeev Khudanpur$^1$ \thanks{\scriptsize{$^\dagger$ Work done during internship at Meta. Contact: huangruizhe09@gmail.com}}
\end{tabular}
}

\address{$^1$Johns Hopkins University, USA $^2$Meta AI, USA, $^3$Carnegie Melon University, USA, \\ $^4$Xiaomi Corp., Beijing, China, $^5$ Boston University, USA}

\begin{document}
\ninept
\maketitle
\begin{abstract}
      Connectionist temporal classification (CTC) models are known to have peaky output distributions. Such behavior is not a problem for automatic speech recognition (ASR), but it can cause inaccurate forced alignments (FA), especially at finer granularity, \emph{e.g.}, phoneme level. This paper aims at alleviating the peaky behavior for CTC and improve its suitability for forced alignment generation, by leveraging label priors, so that the scores of alignment paths containing fewer blanks are boosted and maximized during training. As a result, our CTC model produces less peaky posteriors and is able to more accurately predict the offset of the tokens besides their onset. It outperforms the standard CTC model and a heuristics-based approach for obtaining CTC’s token offset timestamps by $12-40\%$ in phoneme and word boundary errors (PBE and WBE) measured on the Buckeye and TIMIT data. Compared with the most widely used FA toolkit Montreal Forced Aligner (MFA), our method performs similarly on PBE/WBE on Buckeye, yet falls behind MFA on TIMIT. Nevertheless, our method has a much simpler training pipeline and better runtime efficiency. Our training recipe and pretrained model are released in TorchAudio.
\end{abstract}


%
\begin{keywords}
CTC, forced alignment, label priors
\end{keywords}

\vspace{-4pt}
\section{Introduction}
\label{sec:intro}
\vspace{-4pt}

Speech-to-text forced alignment (FA) is a task to automatically produce exact time intervals for the written tokens ({\em e.g.}, phonemes, words) in speech recordings. It's a fundamental step in speech dataset preparation~\cite{Pratap2023ScalingST, Le2023VoiceboxTM}, keyword search~\cite{Huang2023BuildingKS}, closed captioning~\cite{Huang2003AutomaticCC},  and analytical tasks such as phonetic and linguistic studies~\cite{Mackenzie2020AssessingTA, Wu2023UsingAF, yuan2018using}.

Traditionally, FA was based on Gaussian Mixture Models (GMM). Popular toolkits include Montreal forced aligner (MFA)~\cite{McAuliffe2017MontrealFA}, FAVE~\cite{Rosenfelder2014FAVEA}, Prosodylab-Aligner~\cite{Gorman2011ProsodylabalignerAT} and Gentle~\cite{Gentle}.
Among them, MFA is the most widely used and considered as state-of-the-art~\cite{Li2022NeufaNN}. However, compared with deep learning based approaches, GMM-based approaches have limited learning capacity and noise robustness, and their multi-stage modeling pipeline is quite complicated.

As deep learning based ASR becomes popular, FA solutions based on neural networks trained with Connectionist Temporal Classification (CTC)~\cite{Zeyer2017CTCIT, Kurzinger2020CTCSegmentationOL, Tian2022BayesRC, bain2022whisperx} 
, attention mechanism~\cite{Li2022NeufaNN, radford2023robust}, Hidden Markov Model (HMM) topology~\cite{Zhang2021OnLB} or self/semi-supervised learning~\cite{Zhu2021PhonetoAudioAW} have emerged. Among those, CTC-based FA is the most popular option, as it's the by-product of CTC-based ASR and it's very easy to train. 

\begin{figure}[t!]
  \centering
  \includegraphics[width=1.0\linewidth]{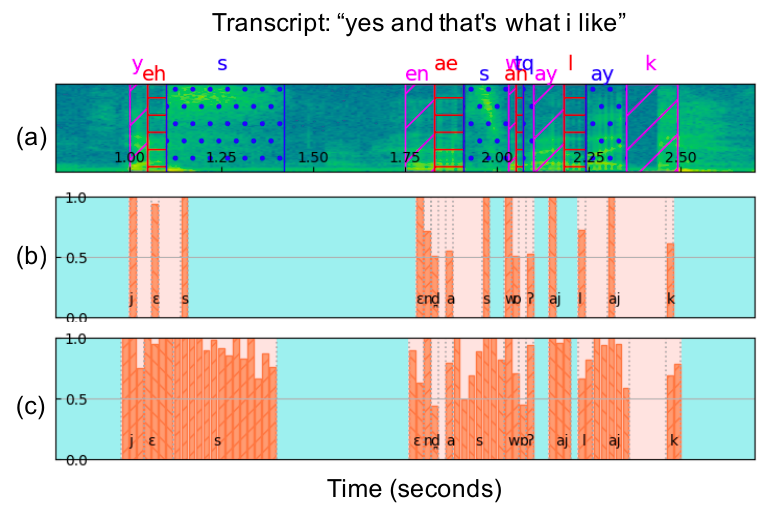}
  \vspace{-24pt}
  \caption{(a) Spectrogram and human-labeled phoneme-level timestamps from Buckeye; (b) Posteriors of the best alignment path of a standard CTC model, with a peaky behavior where each symbol fires for only one frame; (c) Posteriors of the best alignment path of the CTC model trained by our method, where the peaky behavior has been alleviated.}
  \label{fig:less-peaky}
  \vspace{-12pt}
\end{figure}

The major issue with CTC-based FA is its peaky behavior. Originally, CTC was proposed for ASR tasks~\cite{Graves2006ConnectionistTC}. The blank symbol was introduced to serve as a silence token and a placeholder for aligning features/token sequences of different lengths. Empirically, people observed that blanks dominate the predicted sequence \cite{Liu2018ConnectionistTC, Li2020ReinterpretingCT, Zeyer2021WhyDC}, {\em i.e.}, CTC models tend to output spikes of non-blank symbols surrounded by many blanks (Figure~\ref{fig:less-peaky}.(b)). This is not a problem for ASR, which concerns only the accuracy of hypothesis with blanks removed. However, FA task needs to assign token labels to each acoustic frame, such peaky behavior causes inaccurate alignments by assigning too many blanks to non-silence acoustic frames.

To remedy this issue, \cite{Liu2018ConnectionistTC} proposed maximum entropy regularization on sequence level to encourage exploration of paths containing fewer blanks. \cite{Li2020ReinterpretingCT} introduced a strategy that sets the blank and non-blank proportions in the posteriors and focused on key frames during training. \cite{Zeyer2021WhyDC} extended the CTC loss by including label priors, which is known as hybrid model loss. \cite{Variani2018SampledCT} sampled one path from all feasible alignments and converted CTC to cross entropy (CE) training. \cite{Huang2023BuildingKS} used a simple heuristic to smooth the CTC posteriors where each token is assumed to last for a constant number of frames until the next non-blank token is predicted. 


Compared to previous work, this paper takes \cite{Zeyer2021WhyDC} further by applying label priors on real-world FA tasks instead of toy examples (For a comparison with concurrent work~\cite{Chen2023ImprovingFC}, see the appendix). We show that with this simple yet effective method, CTC's peaky behavior can be alleviated and and the FA accuracy is improved (Fig~\ref{fig:less-peaky}.(c)). We apply the CTC loss with label priors on various model architectures, modeling units and model downsampling rates. It turns out a small TDNN-FFN model with $5M$ parameters works the best, as opposed to ASR where Conformer~\cite{gulati2020conformer} is considered better. We derive the gradients of CTC loss with label priors to understand the role of the label priors in the optimization process. Finally, we provide the first experimental benchmark of deep learning based FA methods against GMM-based FA methods {\em on data with human transcribed phoneme-level timestamps}\footnote{Previous work~\cite{Kurzinger2020CTCSegmentationOL} evaluates only on sentence-level. \cite{Li2022NeufaNN} reported better performance than MFA on Buckeye. Yet their method requires very strong supervision (human annotated phoneme boundaries), which is impractical.}. We show that our method significantly improves alignment accuracy over the standard CTC model and a heuristic-based approach~\cite{Huang2023BuildingKS} for better predicting CTC’s token on-/offset. Moreover, we rival the state-of-the-art MFA~\cite{McAuliffe2017MontrealFA} on Buckeye data and perform close to MFA on TIMIT. Nonetheless, our method has a simpler pipeline and faster runtime. Our recipe and pretrained model are available in TorchAudio\footnote{\url{https://github.com/pytorch/audio}}~\cite{Hwang2023TorchAudio2A}.

\section{Connectionist temporal classification revisited}
\label{sec:ctc}

\subsection{Standard CTC loss}

Consider a speech feature sequence $\mathbf{X} = (\mathbf{x}_1, \mathbf{x}_2, ..., \mathbf{x}_T)$ of length $T$, where $\mathbf{x}_t \in \mathbb{R}^D$ is a $D$-dimensional feature vector; a corresponding token sequence $\mathbf{\bar{W}} = (\mathbf{w}_1, \mathbf{w}_2, ..., \mathbf{w}_U)$ of length $U$; and a token vocabulary $\mathcal{V}$. The ASR task is to predict the most probable token sequence $\mathbf{\hat{W}} \in \mathcal{V}^*$ given the features $\mathbf{X}$: 
\begin{equation}
 \vspace{-1pt}
 \mathbf{\hat{W}} = \arg \max_{\mathbf{W} \in \mathcal{V}^*} p(\mathbf{W} | \mathbf{X})     
 \vspace{-1pt}
\end{equation}
CTC~\cite{Graves2006ConnectionistTC} is proposed to compute this probability $p(\mathbf{W} | \mathbf{X})$ directly during training. It introduces a blank symbol $\oslash$ to make an extended vocabulary $\mathcal{V}':=\mathcal{V}\cup\{\oslash\}$. The alignment paths are defined as sequences $\pi \in \mathcal{V}'^{T}$ of the same length as $\mathbf{X}$. Then, the token sequence $\mathbf{W}$ can be obtained by merging repeated tokens and removing any blank tokens $\oslash$ in $\pi$. This operation is defined by a many-to-one function: $\mathbf{W} = \mathcal{B}(\pi)$. CTC proposes to compute $p(\mathbf{W} | \mathbf{X})$ by summing over all valid alignment paths by dynamic programming:
\begin{equation} \label{eq:ctc}
  \vspace{-1pt}
  P_{\mathrm{ctc}}(\mathbf{W}|\mathbf{X}) = \sum_{\pi \in \mathcal{B}^{-1}(\mathbf{W})}P(\pi|\mathbf{X})    
  \vspace{-1pt}
\end{equation}
The probability for each alignment $\pi$ is computed under a conditional independence assumption for each time step:
\begin{equation}
  \vspace{-1pt}
  P(\pi|\mathbf{X}) = \prod_{t=1}^{T} y_{\pi_t}^{t}    
  \vspace{-1pt}
\end{equation}
where $y_{\pi_t}^{t}$ is the posterior probability of token ${\pi_t} \in \mathcal{V}'$ at time $t$ predicted by the model given $\mathbf{X}$. Thus, the model can be trained by the principle of
maximum likelihood, {\em i.e.}, maximizing $P_{\mathrm{ctc}}(\mathbf{W}|\mathbf{X})$ over the training data, or equivalently, minimizing $-\log P_{\mathrm{ctc}}(\mathbf{W}|\mathbf{X})$.

\vspace{-6pt}
\subsection{Optimal alignment path selection problem}

Given a model trained by CTC and an input feature sequence $\mathbf{X}$ of length $T$, we can generate a frame-wise posterior distribution (``posteriorgram'') of shape $T \times |\mathcal{V}'|$ along the time axis and over the extended vocabulary. In other words, the model probabilistically associates each frame in $\mathbf{X}$ with either a token in $\mathcal{V}$ or the blank token $\oslash$. At sequence level, we hope to find the optimal alignment path $\pi^{*} \in \mathcal{B}^{-1}(\mathbf{W})$ with the highest probability $P(\pi|\mathbf{X})$ given the transcription $\mathbf{W}$. This is the optimal alignment path selection problem. This problem can be solved in polynomial time with Viterbi algorithm. From the best alignment path, we can obtain the timestamp for each token in the audio. Note, for CTC FA, the alignments are learned without any frame-level supervision, unlike~\cite{Li2022NeufaNN} or~\cite{Zhu2021PhonetoAudioAW}.

\vspace{-6pt}
\subsection{CTC loss with label priors}
\label{sec-label_priors}

It's well-known that standard CTC models have a peaky behavior. In literature~\cite{Liu2018ConnectionistTC, Li2020ReinterpretingCT, Zeyer2021WhyDC}, there are in-depth analyses of why such behavior happens. In summary, since the blank token $\oslash$ is the most versatile and frequent token in the space of feasible alignment paths $\mathcal{B}^{-1}(\mathbf{W}) \subseteq \mathcal{V}'^{T}$, it is easier for the CTC model to pick up optimal paths containing more blanks at the beginning. Once the paths containing blanks are learned, the model can reinforce itself by tending to go through such paths when computing $P_{\mathrm{ctc}}$, getting gradients dominantly for such paths and ignoring alternative alignment paths. In this way, the peaky CTC posteriors come into place.

To address the peaky behavior issue, i.e., reducing the number of blanks in the optimal alignment paths, it is natural to use unigram label priors $P(k)$, $k \in \mathcal{V}'$ to penalize paths containing too many blank symbols. We can use the following loss function in place of the standard CTC loss (Equation~\ref{eq:ctc}):
\begin{align}
  \vspace{-4pt}
  P_{\mathrm{ctc\_with\_priors}}(\mathbf{W}|\mathbf{X}) &= \sum_{\pi \in \mathcal{B}^{-1}(\mathbf{W})}P_{\mathrm{with\_priors}}(\pi|\mathbf{X}), \label{eq:ctc-with-priors1} \\
  \mathrm{where,\;\;} P_{\mathrm{with\_priors}}(\pi|\mathbf{X}) &= \prod_{t=1}^{T} y_{\pi_t}^{t} / P(\pi_t)^{\alpha} \label{eq:ctc-with-priors2}
  \vspace{-4pt}
\end{align}
The hyper-parameter $\alpha \in \mathbb{R}$ is a scaling factor. When $\alpha = 0$, no priors are applied and this is exactly the standard CTC loss. 
Intuitively, in Equation~\ref{eq:ctc-with-priors2}, if a token, {\em e.g.}, the blank token, occurs more frequently, it will have a larger prior $P(y_{\pi_t}^{t})$ and will get more penalty in its posterior probability, so that all alignment paths including the optimal one will avoid such token. Note that the label priors are not only applied during the Viterbi search of optimal paths, but are also applied during training. In this way, the model intrinsically learns to produce paths containing fewer blanks.



In fact, the idea of leveraging label priors originated from hybrid NN-HMM models for ASR~\cite{Zeyer2021WhyDC, Hinton2012DeepNN}, where the output posteriors of neural networks (NN) are divided by label priors before integrating into the generative framework of HMM. To train the NN models, the alignment outputs from GMM-HMM models are used as the supervision for a frame-level cross entropy loss. In contrast, here we train the model with CTC loss directly without frame-level supervision.


During training, label priors $P(k)$ are first initialized to be a uniform distribution. Later, they are updated at the end of every epoch till convergence. We follow~\cite{Manohar2015SemisupervisedMM} to compute $P(k)$ by marginalizing over the posteriorgram for each token in $ \mathcal{V}'$ over all frames, and accumulate the statistics in the training epoch. Alternatively, we can get $P(k)$ by counting token frequency in the Viterbi alignment paths on training examples. It turns out two methods both converge and produce similar results, while the first one is simpler to implement.


\vspace{-5pt}
\subsection{Gradients of CTC loss with label priors}
\label{sec-gradients}

To understand the impact of label priors in the optimization process, we derive the gradients of Equation~\ref{eq:ctc-with-priors1} following the notations in~\cite{Graves2012SupervisedSL}. The objective function $O$ is defined as the negative logarithm of $P_{\mathrm{ctc\_with\_priors}}$. The un-normalized network output is denoted as $u^t_k$ such that $y^t_k = \text{Softmax}(u^t_k)$. The gradients of CTC loss with label priors is in Equation~\ref{eq-gradents} below, which has a very simple form and is reminiscent of the gradients without label priors as in~\cite{Graves2012SupervisedSL}:
\begin{equation} 
\label{eq-gradents}
\frac{\partial O_{\mathrm{with\_priors}}}{\partial u_k^t} = y_k^t - \frac{1}{p^{\star}(\mathbf{\bar{W}}|\mathbf{X})}\sum_{s \in lab(\mathbf{\bar{W}},k))}\alpha_{t}^{\star}(s)\beta_{t}^{\star}(s)
\end{equation}
To save space, we use the star ($\star$) symbol to denote ``$\mathrm{\_with\_priors}$''. Basically, the gradient of the objective $O$ with respect to the un-normalized network output $u^t_k$ for symbol $k$ at time $t$ consists of two terms. The first term $y_k^t$ is the posterior probability without applying label priors, which is exactly the same as the corresponding term in the gradients in~\cite{Graves2012SupervisedSL}. The only difference is in the second term, with or without label priors. In Equation~\ref{eq-gradents}, it computes how much proportion of $p^{\star}(\mathbf{\bar{W}}|\mathbf{X})$ go through the symbol $k$ at time $t$ after the label priors are applied. The optimization process just tries to match posterior $y_k^t$ with such proportion. If the two terms are equal, the gradient becomes zero and an local optimum is reached. This elaborates the ``error signals'' received by the network during training.

\section{Experiments And Analysis}
\label{sec:exp}

\subsection{Datasets}

We train our model using Librispeech~\cite{Panayotov2015LibrispeechAA} of 960 hours of read English speech from audio books. For evaluation, we use Buckeye~\cite{Pitt2005TheBC} and TIMIT~\cite{garofolo1993timit} corpora. TIMIT contains 5.4 hours of read speech with time-aligned phoneme-level transcriptions. Buckeye contains spontaneous English conversations (interviews) of 40 speakers and 20 hours. It comes with forced aligned then manually corrected phoneme-level timestamps. Buckeye comes in many 10-minute-ish recordings. Following the common practice in~\cite{McAuliffe2017MontrealFA, Li2022NeufaNN}, we segment the long recordings into smaller chunks. We take chunks separated by non-speech (pauses, noise, etc.) of more than one second instead of 150 ms as in \cite{McAuliffe2017MontrealFA} to avoid too short segments. 

We randomly selected 4 speakers as development data to tune hyper parameters. Note, however, for the FA task, it is always possible to fine-tune the model on the target/test data (audio and transcription) to reduce acoustic condition mismatch, before actually producing alignments. We will provide FA results with and without fine-tuning on the target data.

\subsection{Metrics}

Phoneme or word boundary error (PBE/WBE) is to measure how close the predicted and manually labeled timestamps are. Following~\cite{McAuliffe2017MontrealFA}, PBE is defined as the average of $N$ utterance-level PBEs (where $p$ stands for phonemes and $u_i$ is the i-th utterance below): 
\small
\begin{equation*}
    PBE \doteq \frac{1}{N} \sum_i \frac{1}{|u_i|} \sum_{p \in u_i} \frac{1}{2} \left ( |p_{beg}^{ref} - p_{beg}^{pred}| + |p_{end}^{ref} - p_{end}^{pred}| \right )
\end{equation*}
\normalsize
WBE is similarly defined on word-level. Ideally, PBE and WBE should be close to 0. 

Phoneme or word average duration (PDUR/WDUR) is the average predicted duration of phonemes or words. PDUR and WDUR should match the average duration of manual timestamps, the closer the better. PDUR, WDUR and the final label priors are used to measure the spikiness of CTC models.


\subsection{Model configuration and implementation details}

For the encoder of our CTC aligner, we compared different model architectures (TDNN-FFN, TDNN-BLSTM~\cite{Zhang2021OnLB} and Conformer~\cite{gulati2020conformer} of $5M$, $27M$ and $85M$ parameters, respectively), different modeling units (phonemes, characters, sentencepieces~\cite{Kudo2018SentencePieceAS}) and different sub-sampling rates ($1$, $2$ and $4$). Our models take Mel-spectrogram features with 10 ms frame shift as input, and are trained by minimizing the CTC loss with or without label priors (Section~\ref{sec-label_priors}). For phoneme models, the phoneme set (of $93$ phonemes), pronunciation dictionary and G2P model are taken from MFA~\cite{mfa_english_mfa_dictionary_2022}. 
We train our models for $20$ epochs and select the model with the best loss value on development data. We observed no further improvement for FA when training for more epochs, which is different from ASR.


Noteworthy is that, at the time we work on this paper, the CTC loss implementation~\footnote{\tiny\url{https://pytorch.org/docs/stable/generated/torch.nn.CTCLoss.html}} in Pytorch only supports inputs that are sum-to-one probabilities. However, with label priors, the inputs to CTC do {\em not} sum to one anymore. On the other hand, the CTC dynamic programming does not require a valid probability distribution. As alternatives for Pytorch's CTC, we found the CTC loss implemented in k2~\footnote{\tiny\url{https://k2-fsa.github.io/k2/python_api/api.html\#ctc-loss}} library or in another open source implementation~\footnote{\tiny\url{https://github.com/vadimkantorov/ctc}} can match the gradients computed by Equation~\ref{eq-gradents}, when label priors are applied.

\vspace{-4pt}
\subsection{Results}

\subsubsection{Effectiveness of the proposed CTC model}

We trained a TDNN-FFN model of $5M$ parameters with phoneme outputs, which is a stack of 3 TDNN layers (with kernel size $5, 3, 3$ and stride (sub-sampling factor of the acoustic encoder) size $2, 1, 1$) and 5 feedforward layers. In Table~\ref{tab:table1}, we compare our aligner with the aligner trained with standard CTC loss, the aligners based on Wav2Vec2~\cite{Baevski2020wav2vec2A} \footnote{\tiny\url{https://pytorch.org/audio/main/tutorials/forced_alignment_tutorial.html}} or MMS \cite{Pratap2023ScalingST} \footnote{\tiny\url{https://pytorch.org/audio/main/tutorials/ctc_forced_alignment_api_tutorial.html}} models, heuristics~\cite{Huang2023BuildingKS} as well as MFA based on GMM-HMM triphone model.

\vspace{-10pt}
\begin{table}[h!]
\centering
\caption{Comparing the alignment accuracy of several FA solutions on Buckeye and TIMIT data. All metrics are in milliseconds. The closer to the ground truth (the last row), the better.}
\label{tab:table1}
\resizebox{\columnwidth}{!}{%
\begin{tabular}{|cc|cccccccc}
\hline
\multicolumn{2}{|c|}{}                           & \multicolumn{4}{c|}{Buckeye}                                                                                                                                                                & \multicolumn{4}{c|}{TIMIT}                                                                                                                                                                  \\ \cline{3-10}
\multicolumn{2}{|c|}{\multirow{-2}{*}{}}         & \multicolumn{1}{c|}{\rotatebox{70}{PBE}}         & \multicolumn{1}{c|}{\rotatebox{70}{WBE}}         & \multicolumn{1}{c|}{\cellcolor[HTML]{E6E5E5}\rotatebox{70}{PDUR}}        & \multicolumn{1}{c|}{\cellcolor[HTML]{E6E5E5}\rotatebox{70}{WDUR}}         & \multicolumn{1}{c|}{\rotatebox{70}{PBE}}         & \multicolumn{1}{c|}{\rotatebox{70}{WBE}}         & \multicolumn{1}{c|}{\cellcolor[HTML]{E6E5E5}\rotatebox{70}{PDUR}}        & \multicolumn{1}{c|}{\cellcolor[HTML]{E6E5E5}\rotatebox{70}{WDUR}}         \\ \hline
\multicolumn{1}{|c|}{1}  & Wav2Vec2~\cite{Baevski2020wav2vec2A}              & \multicolumn{1}{c|}{-}           & \multicolumn{1}{c|}{89}          & \multicolumn{1}{c|}{\cellcolor[HTML]{E6E5E5}-}           & \multicolumn{1}{c|}{\cellcolor[HTML]{E6E5E5}177}          & \multicolumn{1}{c|}{-}           & \multicolumn{1}{c|}{48}          & \multicolumn{1}{c|}{\cellcolor[HTML]{E6E5E5}-}           & \multicolumn{1}{c|}{\cellcolor[HTML]{E6E5E5}229}          \\
\multicolumn{1}{|c|}{2}  & MMS~\cite{Pratap2023ScalingST}                   & \multicolumn{1}{c|}{-}           & \multicolumn{1}{c|}{53}          & \multicolumn{1}{c|}{\cellcolor[HTML]{E6E5E5}-}           & \multicolumn{1}{c|}{\cellcolor[HTML]{E6E5E5}181}          & \multicolumn{1}{c|}{-}           & \multicolumn{1}{c|}{37}          & \multicolumn{1}{c|}{\cellcolor[HTML]{E6E5E5}-}           & \multicolumn{1}{c|}{\cellcolor[HTML]{E6E5E5}242}          \\
\multicolumn{1}{|c|}{3}  & Heuristics~\cite{Huang2023BuildingKS}            & \multicolumn{1}{c|}{42}          & \multicolumn{1}{c|}{55}          & \multicolumn{1}{c|}{\cellcolor[HTML]{E6E5E5}60}          & \multicolumn{1}{c|}{\cellcolor[HTML]{E6E5E5}212}          & \multicolumn{1}{c|}{31}          & \multicolumn{1}{c|}{41}          & \multicolumn{1}{c|}{\cellcolor[HTML]{E6E5E5}30}          & \multicolumn{1}{c|}{\cellcolor[HTML]{E6E5E5}239}          \\
\multicolumn{1}{|c|}{4}  & MFA~\cite{McAuliffe2017MontrealFA}                   & \multicolumn{1}{c|}{30}          & \multicolumn{1}{c|}{41}          & \multicolumn{1}{c|}{\cellcolor[HTML]{E6E5E5}84}          & \multicolumn{1}{c|}{\cellcolor[HTML]{E6E5E5}251}          & \multicolumn{1}{c|}{17}          & \multicolumn{1}{c|}{23}          & \multicolumn{1}{c|}{\cellcolor[HTML]{E6E5E5}85}          & \multicolumn{1}{c|}{\cellcolor[HTML]{E6E5E5}313}          \\
\multicolumn{1}{|c|}{5}  & \textbf{+fine-tuning} & \multicolumn{1}{c|}{\textbf{27}} & \multicolumn{1}{c|}{\textbf{36}} & \multicolumn{1}{c|}{\cellcolor[HTML]{E6E5E5}\textbf{83}} & \multicolumn{1}{c|}{\cellcolor[HTML]{E6E5E5}\textbf{250}} & \multicolumn{1}{c|}{\textbf{16}} & \multicolumn{1}{c|}{\textbf{22}} & \multicolumn{1}{c|}{\cellcolor[HTML]{E6E5E5}\textbf{86}} & \multicolumn{1}{c|}{\cellcolor[HTML]{E6E5E5}\textbf{314}} \\ \hline
\multicolumn{1}{|c|}{6}  & Standard CTC          & \multicolumn{1}{c|}{44}          & \multicolumn{1}{c|}{58}          & \multicolumn{1}{c|}{\cellcolor[HTML]{E6E5E5}21}          & \multicolumn{1}{c|}{\cellcolor[HTML]{E6E5E5}169}          & \multicolumn{1}{c|}{32}          & \multicolumn{1}{c|}{42}          & \multicolumn{1}{c|}{\cellcolor[HTML]{E6E5E5}21}          & \multicolumn{1}{c|}{\cellcolor[HTML]{E6E5E5}229}          \\
\multicolumn{1}{|c|}{7}  & +fine-tuning          & \multicolumn{1}{c|}{39}          & \multicolumn{1}{c|}{52}          & \multicolumn{1}{c|}{\cellcolor[HTML]{E6E5E5}22}          & \multicolumn{1}{c|}{\cellcolor[HTML]{E6E5E5}163}          & \multicolumn{1}{c|}{31}          & \multicolumn{1}{c|}{40}          & \multicolumn{1}{c|}{\cellcolor[HTML]{E6E5E5}23}          & \multicolumn{1}{c|}{\cellcolor[HTML]{E6E5E5}234}          \\ \hline
\multicolumn{1}{|c|}{8}  & Our CTC               & \multicolumn{1}{c|}{38}          & \multicolumn{1}{c|}{43}          & \multicolumn{1}{c|}{\cellcolor[HTML]{E6E5E5}64}          & \multicolumn{1}{c|}{\cellcolor[HTML]{E6E5E5}221}          & \multicolumn{1}{c|}{28}          & \multicolumn{1}{c|}{29}          & \multicolumn{1}{c|}{\cellcolor[HTML]{E6E5E5}72}          & \multicolumn{1}{c|}{\cellcolor[HTML]{E6E5E5}288}          \\
\multicolumn{1}{|c|}{9}  & \textbf{+fine-tuning} & \multicolumn{1}{c|}{\textbf{30}} & \multicolumn{1}{c|}{\textbf{34}} & \multicolumn{1}{c|}{\cellcolor[HTML]{E6E5E5}\textbf{74}} & \multicolumn{1}{c|}{\cellcolor[HTML]{E6E5E5}\textbf{232}} & \multicolumn{1}{c|}{\textbf{27}} & \multicolumn{1}{c|}{\textbf{28}} & \multicolumn{1}{c|}{\cellcolor[HTML]{E6E5E5}\textbf{79}} & \multicolumn{1}{c|}{\cellcolor[HTML]{E6E5E5}\textbf{301}} \\ \hline
\multicolumn{1}{|c|}{10} & Ground truth          & \multicolumn{1}{c|}{0}           & \multicolumn{1}{c|}{0}           & \multicolumn{1}{c|}{\cellcolor[HTML]{E6E5E5}82}          & \multicolumn{1}{c|}{\cellcolor[HTML]{E6E5E5}241}          & \multicolumn{1}{c|}{0}           & \multicolumn{1}{c|}{0}           & \multicolumn{1}{c|}{\cellcolor[HTML]{E6E5E5}76}          & \multicolumn{1}{c|}{\cellcolor[HTML]{E6E5E5}305}          \\ \hline
\end{tabular}%
}
\end{table}


From Table~\ref{tab:table1}, the CTC model trained by our proposed method (row 8, $\alpha=0.3$\footnote{\tiny When $\alpha=0$, the results are exactly the same as row 6; when $\alpha>0.3$, the model does not converge.}) clearly outperforms the standard CTC model (row 6), the heuristic-based method (row 3) as well as Wav2Vec2/MMS aligners (row 1, 2) in all metrics for the FA task. In particular, the predicted PDUR value being $21$ ms in row 6 matches the input frame size of the model ($20$ ms with stride size $2$), which means the non-blank tokens fire for just one frame. In contrast, our model's PDUR (row 8) is closer to the ground truth (row 10). On the other hand, the label prior probability for the blank token on training data are $0.80$ and $0.32$ for row 6 and 8 respectively (which is not shown in the table). This verifies that the standard CTC model is indeed peaky, and our proposed method can effectively reduce the peakiness (Fig.~\ref{fig:less-peaky}).

Note that the MFA model we used (english\_mfa) is trained on $3700$+ hours of English speech including Librispeech. To reduce acoustic condition mismatch and further improve FA, we finetune MFA and our aligners (for 6 epochs) on Buckeye data. From Table~\ref{tab:table1}, finetuning is effective (row 5, 7, 9 vs. row 4, 6, 8) as all metrics become closer to the ground truth. 
On Buckeye, our best BPE/WBE results (row 9) are close to the best of MFA (row 5), whereas WBE even outperforms MFA slightly. We conjecture that our aligner is good at modeling tokens next to silence ({\em e.g.}, word boundaries), due to the larger modeling capacity of neural networks compared to GMMs. On TIMIT, our best aligner still falls behind MFA. We notice our aligner and MFA have different frame rates, $20$ ms and $10$ ms respectively, which is probably the reason for our aligner not being able to produce even finer-grained timestamps. However, when we set our frame rate to be $10$ ms, the CTC model becomes harder to train due to longer input sequences~\cite{Zeyer2021WhyDC} and does not produce better alignments. This is left as future work.

For PBE/WBE improvement, we investigate it further by breaking down PBE/WBE (of the models in Table~\ref{tab:table1} row 6 and 8) into onset/offset timestamps errors. From Table~\ref{tab:table4}, we find that the standard CTC makes more errors on onset prediction than offset prediction, which aligns with our expectation that a standard CTC model tends to delay token predictions. On the other hand, our CTC model improves {\em both} onset and offset predictions, with more improvements on the onset side, so that the onset/offset errors are more balanced.  


\vspace{-10pt}
\begin{table}[h!]
\centering
\caption{The breakdown of onset/offset timestamps errors on Buckeye for Table~\ref{tab:table1} row 6 and 8. All metrics are in milliseconds. }
\label{tab:table4}
\resizebox{0.7\columnwidth}{!}{%
\begin{tabular}{|c|cc|cc|}
\hline
\multirow{2}{*}{} & \multicolumn{2}{c|}{phoneme} & \multicolumn{2}{c|}{word} \\ \cline{2-5} 
 & \multicolumn{1}{c|}{onset} & offset & \multicolumn{1}{c|}{onset} & offset \\ \hline
Standard CTC & \multicolumn{1}{c|}{51} & 39 & \multicolumn{1}{c|}{63} & 54 \\
Our CTC & \multicolumn{1}{c|}{\textbf{39}} & \textbf{36} & \multicolumn{1}{c|}{\textbf{44}} & \textbf{42} \\ \hline
\end{tabular}
}
\vspace{-4pt}
\end{table}

In above experiments, it takes MFA $9.5$ minutes to finish alignment generation on $20$-hour Buckeye data, with $8$ CPU jobs and multithreading turned on. In contrast, it takes only $3$ minutes for our model to finish on $1$ NVIDIA Titan RTX GPU, or less than $1$ minute on $4$ GPUs, thanks to the PyTorch-based implementation. This is not an apple-to-apple comparison given MFA does not support GPU, yet it gives us a rough estimate of the best runtime of both systems. 

\vspace{-8pt}
\subsubsection{Impact of network configurations}
\label{configurations}

We start with the baseline configuration (TDNN-FFN, stride=2, phoneme) as in Table~\ref{tab:table2} row 1 (corresponding to row 6 and 8 in Table~\ref{tab:table1}), and vary the model architecture (row 2-3), modeling unit (row 4-5), and model stride size (row 6-7) independently on top of the baseline configuration. We report results on Buckeye {\em with and without} applying our method in each cell of Table~\ref{tab:table2}.

\begin{table}[t!]
\caption{Comparing different network configurations on Buckeye. The baseline configuration is (TDNN-FFN, stride=2, phoneme). We vary the model architecture, modeling unit and model stride size independently on top of the baseline configuration. In each cell, results for the \textbf{standard/proposed} CTC model are reported. All metrics are in milliseconds. The closer to ground truth (the last row), the better.}
\label{tab:table2}
\centering
\small
\resizebox{0.85\columnwidth}{!}{%
\begin{tabular}{|cc|c|c|
>{\columncolor[HTML]{E6E5E5}}c |
>{\columncolor[HTML]{E6E5E5}}c |}
\hline
\multicolumn{2}{|c|}{}                  & PBE     & WBE      & PDUR    & WDUR      \\ \hline
\multicolumn{1}{|c|}{1} & Baseline      & 44 / \textbf{38} & 58 / \textbf{43}  & 21 / \textbf{64} & 169 / \textbf{221} \\ \hline
\multicolumn{1}{|c|}{2} & TDNN-BLSTM    & \textbf{42} / 53 & \textbf{53} / 74  & 23 / \textbf{75} & 175 / \textbf{261} \\
\multicolumn{1}{|c|}{3} & Conformer     & \textbf{43} / 55 & \textbf{51} / 75  & 25 / \textbf{77} & 180 / \textbf{278} \\ \hline
\multicolumn{1}{|c|}{4} & char          & -       & 62 / \textbf{52}  & -       & 196 / \textbf{235} \\
\multicolumn{1}{|c|}{5} & sentencepiece & -       & 101 / \textbf{52} & -       & 80 / \textbf{230}  \\ \hline
\multicolumn{1}{|c|}{6} & stride=1      & 51 / \textbf{46} & 65 / \textbf{49}  & 11 / \textbf{78} & 156 / \textbf{242} \\
\multicolumn{1}{|c|}{7} & stride=4      & 45 / \textbf{40} & 56 / \textbf{47}  & 43 / \textbf{71} & 196 / \textbf{230} \\ \hline
\multicolumn{1}{|c|}{8} & Ground truth  & 0       & 0        & 82      & 241       \\ \hline
\end{tabular}%
}
\vspace{-8pt}
\end{table}

When the models with long-range memory are used (Table~\ref{tab:table2}, row 2 and 3), our method actually degrades PBE/WBE. It turns out such models have learned to predict non-blank tokens repeatedly even when there is no speech in the audio, in order to avoid blank penalties from the priors. Only the TDNN-FFN model with $5M$ parameters works well, which is quite different from ASR where Conformer is considered better. Note, our TDNN-FFN model has a very limited perception range over the input features (13 frames), so it only has the ability to learn the short-range acoustic information, instead of the long-range language dependencies. In other words, $5M$ parameters are probably sufficient to make a good acoustic model. Thus, we conjecture that in the $85M$-param Conformer ASR model, only a small portion of parameters are used for acoustic modeling, whereas the rest of parameters are used for language modeling of long-range dependency. 




As shown in rows 4$\sim$7, the proposed CTC (right) works better than standard CTC (left) in each cell, but none of the other configurations in rows 4$\sim$7 work better than row 1. From the PDUR column, the standard CTC model (left) tends to predict PDUR to be almost the same size as the model input frame size, while PDUR predicted by our CTC model (right) is closer to the ground truth.

\vspace{-5pt}
\subsubsection{Other variations of the proposed method}

We experimented with the following variations of the proposed CTC model: (1) we use label priors only during decoding the standard CTC model, and got no improvements, which suggests incorporating priors into training loss is important. (2) Instead of applying label priors to all tokens, we apply penalties only to the blank tokens to the standard CTC model during training, and got only minimal improvements over standard CTC. (3) We disallow intra-word blanks during alignment for Table~\ref{tab:table1} row 8 or 9, which resembles 1-state HMM topology~\cite{Zhang2021OnLB} disallowing intra-word blanks, and got minimal improvements over the standard CTC. This result agrees with Figure. 3 from \cite{Zeyer2021WhyDC}, where models with HMM topology tend to clump up non-blank tokens.
(4) We fine-tune the standard CTC model with the proposed loss on Buckeye, and got similar results as our best results, suggesting that fine-tuning a standard CTC model with the proposed method is sufficient to resolve the peaky behavior.

\vspace{-6pt}
\section{Conclusion and Future work}
\label{sec:conclusion}
\vspace{-2pt}

In this paper, we proposed using a CTC loss with label priors to train CTC models to produce less peaky distribution, enabling the CTC model to more accurately predict not only the onset but also the offset of the tokens. Such property is especially suitable for the forced alignment (FA) task. From our benchmarks using human transcribed phoneme-level timestamps, we see both the predicted onset and offset timestamps are improved by the proposed model, leading to significant alignment accuracy improvement (measured by phoneme/word boundary errors and phoneme/word durations) compared to the standard CTC model and a heuristics-based baseline method. Our proposed model also rivals the performance of the state-of-the-art MFA aligner on Buckeye and TIMIT data.

On top of this work, there are several directions to further improve the CTC-based or deep learning based forced aligners. 
Firstly, the neural aligners' noise robustness to either noisy audios or noisy transcriptions should be investigated. Neural networks are considered more robust than GMMs.
Second, besides the light-weighted TDNN-FFN model, it would be useful if the peaky behavior of the TDNN-BLSTM or Conformer based CTC models can also be reduced. In this way, we can re-purpose pretrained ASR systems, which are usually based on Conformer-type architectures nowadays, for FA applications.
Third, methods to effectively train FA models with smaller sub-sampling rates should be investigated. Finally, it's worthwhile to explore if self-supervised learning features or objectives can further benefit our proposed CTC aligner.



\appendix
\section{Appendix}
There is an independent concurrent work~\cite{Chen2023ImprovingFC} which is derived from the same theoretical foundation~\cite{Zeyer2021WhyDC} as ours. We briefly summarize the contributions and main differences.

Both this paper and~\cite{Chen2023ImprovingFC} adopts label priors to reduce the peakiness of CTC models on real-world FA tasks and to improve FA time stamp accuracy. Both works show applying label priors on different real-world datasets is effective. Notably, both papers' frame-level classifiers are models {\em without} long-range memory: FFN is used in~\cite{Chen2023ImprovingFC} on top of encoder's outputs and speech features, while TDNN-FFN is used in our paper. This validates our choice of model architecture in Section~\ref{configurations}.

In~\cite{Chen2023ImprovingFC}, they derive time stamps for the attention-based encoder-decoder architecture (LAS), while our work is primarily for CTC models. Secondly, ~\cite{Chen2023ImprovingFC} proposes other loss function terms and regularization techniques to improve FA, while our work focuses primarily on label priors and investigate the gradients. Third,~\cite{Chen2023ImprovingFC} employs Tensorflow, Lingvo and RETURNN toolkits (check out their paper for details) for implementation, while our solution is based on Pytorch and is publicly available~\footnote{\tiny{\url{https://github.com/huangruizhe/audio/blob/aligner_label_priors/examples/asr/librispeech_alignment/loss.py}}}. We also spot an issue~\footnote{\tiny{\url{https://github.com/pytorch/pytorch/issues/122243}}} of Pytorch's native CTC implementation and have addressed the issue in our solution. Finally,~\cite{Chen2023ImprovingFC} uses Librispeech (whose ground-truth timetamps are automatically derived by a HMM-GMM model) and internal data for experiments. This work uses TIMIT and Buckeye, which comes with manually annotated timestamps which are fairer when comparing CTC models with HMM-GMM models.

\bibliographystyle{IEEEbib}
\bibliography{strings,refs}



\end{document}